\documentclass[preprint]{aastex631}

\submitjournal{PSJ}
\shorttitle{}
\shortauthors{}

\graphicspath{{./}{figures/}}

\begin{document}

\title{The effect of salinity on ocean circulation and ice-ocean interaction on Enceladus}

\correspondingauthor{Yaoxuan Zeng}
\email{yxzeng@uchicago.edu}

\author[0000-0002-2624-8579]{Yaoxuan Zeng}
\affiliation{Department of the Geophysical Sciences, The University of Chicago, Chicago, IL 60637, USA}

\author[0000-0002-6479-8651]{Malte F. Jansen}
\affiliation{Department of the Geophysical Sciences, The University of Chicago, Chicago, IL 60637, USA}

\begin{abstract}

Observational data suggest that the ice shell on Enceladus is thicker at the equator than at the pole, indicating an equator-to-pole ice flow. If the ice shell is in an equilibrium state, the mass transport of the ice flow must be balanced by the freezing and melting of the ice shell, which in turn is modulated by the ocean heat transport. Here we use a numerical ocean model to study the ice-ocean interaction and ocean circulation on Enceladus with different salinities. We find that salinity fundamentally determines the ocean stratification. A stratified layer forms in the low salinity ocean, affecting the ocean circulation and heat transport. However, in the absence of tidal heating in the ice shell, the ocean heat transport is found to always be towards lower latitudes, resulting in freezing at the poles, which cannot maintain the ice shell geometry against the equator-to-pole ice flow. The simulation results suggest that either the ice shell on Enceladus is not in an equilibrium state, or tidal dissipation in the ice shell is important in maintaining the ice shell geometry. The simulations also suggest that a positive feedback between cross-equatorial ocean heat transport and ice melting results in spontaneous symmetry breaking between the two hemispheres. This feedback may play a role in the observed interhemispheric asymmetry in the ice shell.

\end{abstract}

\keywords{Enceladus -- salinity -- stratified layer -- ocean heat transport -- ice geometry -- symmetry breaking}
\section{Introduction} \label{sec:intro}

Enceladus is believed to have a global ocean of about 40~km in depth under a global ice shell \citep[e.g.,][]{postberg2011salt, patthoff2011fracture, thomas2016enceladus}. The energy source that maintains the global liquid ocean and a heat loss of about 10~GW at the south polar region \citep{spencer2006cassini,howett2011high,spencer2013enceladus} is believed to primarily come from tidal dissipation. In general, tidal dissipation is expected to occur in the ice shell, in the ocean, and in the solid core. Tidal heating in the ocean has been thought to be negligible compared to that in the ice shell and inner solid core \citep[e.g.,][]{chen2011obliquity,tyler2011tidal,chen2014tidal,beuthe2016ocean,hay2017numerically,matsuyama2018ocean,hay2019nonlinear}. Although recent work suggests dissipation of internal tides may generate more heat in a stratified ocean, if resonances occur \citep{rovira2023thin}, the stratification in Enceladus's ocean is unlikely to be strong enough to support a heat generation rate that is comparable to the heat loss rate \citep[e.g.,][]{kang2022saltice}. Tidal dissipation in the ice shell, as well as the turbulent dissipation of water in conduits at the south pole \citep{kite2016sustained}, has been argued to potentially reach O(10~GW) \citep{beuthe2019enceladus,souvcek2019tidal}. However, the rheology of ice on Enceladus is poorly understood, which strongly affects the tidal dissipation rate in the ice shell. The low density of Enceladus suggests that the core is porous, which possibly supports strong tidal dissipation in the core, reaching O(10~GW) \citep{iess2014gravity,roberts2015fluffy,vcadek2016enceladus,choblet2017powering}. Libration has been argued to cause dissipative heating in the ocean of up to O(0.1~GW) \citep{lemasquerier2017libration,wilson2018can,rekier2019internal,soderlund2020ice}. Although the fraction of tidal heating coming from the solid core and the ice shell remains highly uncertain, some heating from the bottom solid core is expected, which fundamentally shapes the ocean circulation on Enceladus.

\setcounter{footnote}{1}

The ice shell on Enceladus is about 20~km thick on average, and is not flat: it is thickest at the equator (over 30~km) and thinnest at the south polar region (less than 10~km). There is also an interhemispheric asymmetry in the ice shell with ice being thinner at the south pole than at the north pole \citep{beuthe2016ice,vcadek2019long,hemingway2019enceladus}. The significant gradient in the ice shell thickness suggests an equator-to-pole ice flow. If the ice shell on Enceladus is in an equilibrium state, the equator-to-pole mass transport of the ice flow must be balanced by ice freezing at the equator and melting at the pole. Ice shell models generally predict that a freezing/melting rate of a few millimeters per year\footnote{Throughout this manuscript we will use years (and similarly months and days) as referring to Earth-years.} on Enceladus is required to balance the equator-to-pole ice flow, or equivalently, O(0.01~W~m$^{-2}$) of latent heat  \citep{vcadek2019long,kang2020ice,kang2022saltice}.

Freezing and melting of the ice is determined by the heat budget. The associated latent heat ($Q_{LH}$, with melting defined as positive) has to match the imbalance between the heat flux from the ocean to the ice ($Q_{oi}$) and the upward heat flux within the ice shell. Since the ice shell on Enceladus is unlikely to be convecting \citep{nimmo2018thermal}, we assume that the upward heat flux in the ice is purely due to conduction ($Q_{cond}$). The energy balance at the ice-ocean interface is then: $Q_{cond} + Q_{LH} = Q_{oi}$. The conductive heat flux at the ice-ocean interface in turn is modulated by the ice shell thickness, as well as the tidal heating in the ice shell ($Q_i$). Without strongly localized tidal heating in the ice shell, conductive heat loss is strongest at the pole where the ice is thinnest, which would favor ice freezing at the pole - the opposite of what is needed to maintain the ice geometry. Tidal dissipation in the ice shell could be a possible answer to the problem, since it is strongest at the pole and is amplified where the ice is thin \citep[e.g., ][]{beuthe2018enceladus,beuthe2019enceladus,kang2020ice}, and thus could potentially allow for melting at the pole. Alternatively, the heat flux from the ocean, $Q_{oi}$, could be amplified at the pole. The global ocean heat flux to the ice comes from the tidal heating in the bottom solid core that peaks at the pole \citep{choblet2017powering}, but the spatial pattern is modulated by the ocean circulation, which is the focus of this paper.

Previous studies have explored the global ocean circulation on Enceladus and other icy moons \citep{soderlund2014ocean,travis2015keeping,soderlund2019ocean,amit2020cooling,lobo2020pole,kang2020core,zeng2021ocean,ashkenazy2021dynamic,bire2022exploring,kvorka2022numerical,kang2022sizeice1,kang2022sizeice2,kang2022symmetry,kang2022saltice,jansen2023energetics}. These studies have used a range of different methods (e.g., conceptual models, ocean-only numerical simulations, and ice-ocean coupled numerical simulations), in either 2-D or 3-D domains, with a wide range of parameters and different boundary conditions. However, there have not yet been simulations with fully coupled ice-ocean thermodynamics in a 3-D global domain. The published simulation results vary substantially in their ocean circulation and ocean-ice heat fluxes, likely as a result of the large uncertainties in the external parameters and the difficulties in modeling the global ocean in a realistic parameter regime. The real regime of ocean circulation and heat transport on Enceladus therefore remains uncertain.

One important factor determining the ocean circulation and heat transport is salinity ($S$). Salinity determines the equation of state of water, thus affecting ocean stratification and circulation. Most importantly, if the salinity is lower than a critical value, the thermal expansivity ($\alpha \equiv -(1/\rho)(\partial \rho/\partial T)$ where $\rho$ is density and $T$ is temperature) is negative near the freezing point. As a result, in such a low salinity ocean, a stably stratified layer forms in the upper ocean with bottom heating \citep{melosh2004temperature,zeng2021ocean}. The threshold for the negative thermal expansivity to exist under the pressure found in Enceladus's ocean is around 20~g~kg$^{-1}$ \citep{zeng2021ocean,kang2022saltice}. There are different methods to estimate the salinity of Enceladus's ocean, based on the observations of salt-rich ice grains \citep{postberg2009salt}, the existence of silica nanoparticles \citep{hsu2015ongoing}, using models of boiling water in ice conduits \citep{ingersoll2016controlled}, equilibrated water-rock interaction \citep{zolotov2007oceanic}, ocean geochemistry \citep{glein2018geochemistry}, and ocean heat transport \citep{kang2022saltice}. The estimates vary from study to study, giving a wide range of 2-40~g~kg$^{-1}$. Therefore, both ``high salinity'' ($S>20$~g~kg$^{-1}$) and ``low salinity'' ($S<20$~g~kg$^{-1}$) scenarios should be carefully considered when studying the ocean circulation on Enceladus.

Salinity can also influence the overturning circulation driven by surface density contrasts, and modulate the ocean heat transport, which further affects the ice geometry. \cite{kang2022saltice} model the ocean circulation with prescribed ice thickness and prescribe melting rates at the pole and freezing rates at the equator as required to balance the estimated ice flow induced by the ice thickness gradient. They find that the surface density contrast between the equator and the pole drives an overturning circulation in each hemisphere. The direction of the circulation depends on the mean salinity of the ocean, but the ocean heat transport is found to always be equatorward because the freezing point at the pole (with thinner ice and hence lower pressure) is higher than at the equator (with thicker ice and hence higher pressure), and the heat transport is always from the warmer pole to the colder equator. The equatorward ocean heat transport favors freezing at the pole and melting at the equator and thereby acts to reduce the equator-to-pole ice shell thickness gradient, which is known as the ice pump effect \citep{lewis1986ice}.

The partitioning of tidal heating between the ice shell and the bottom solid core may also affect the ocean heat transport, further influencing the asymmetric ice shell geometry between the two hemispheres. \cite{kang2022symmetry} prescribe an ice geometry with an interhemispheric asymmetry where ice is slightly thinner in the southern hemisphere. They incorporate fully coupled ice-ocean thermodynamics in a 2-D domain with tidal dissipation in both the ice shell and the bottom solid core. They find that when the heating is dominated by tidal dissipation in the ice shell, there is stronger melting at the south pole than the north, and the overturning circulation transports heat to the south, thus enhancing the asymmetry; instead, when the heating is primarily in the core, there is stronger freezing at the south pole than the north, and the overturning transports heat to the north, thus suppressing the asymmetry.

Here we study the effect of salinity on the ocean circulation and ice-ocean interaction on Enceladus, focusing on the ocean heat transport. We carry out 3-D global ocean simulations using the Massachusetts Institute of Technology General Circulation Model (MITgcm) \citep{adcroft2018mitgcm}, with prescribed ice geometry and fully coupled ice-ocean thermodynamics. We exclude tidal heating in the ice to focus specifically on whether ocean heat transport from the solid core can maintain the observed ice topography and the associated pattern of freezing and melting. Section~\ref{sec:model} describes the model configuration and experimental design. Section~\ref{sec:results} introduces the numerical simulation results, focusing on ocean stratification, circulation, and heat transport. Section~\ref{sec:conclusion} compares the simulation results with previous studies and provides concluding remarks.

\begin{deluxetable*}{cc}
\tablenum{1}
\tablecaption{Parameters for Enceladus simulations \label{tb:parameters}}
\tablewidth{0pt}
\tablehead{
\colhead{Parameters} & \colhead{Value}
}
\startdata
Surface gravity $g(r_s)$ & 0.113 m s$^{-2}$ \\
Rotation rate $\Omega$ & $5.31\times 10^{-5}$ s$^{-1}$ \\
Ocean bottom radius $r_c$ & 192 km \\
Mean ice-ocean interface radius $r_o$ & 232 km \\
Mean ice surface radius $r_i$ & 252 km \\
\enddata
\tablecomments{The values are taken from \citet{vcadek2019long} with some simplifications.}
\end{deluxetable*}

\begin{deluxetable*}{ccccc}
\tablenum{2}
\tablecaption{Experimental Design\label{tb:experiments}}
\tablewidth{0pt}
\tablehead{
\colhead{Case} & \colhead{Case} & \colhead{Mean ocean} &  \colhead{Horizontal turbulent} \\
\colhead{label} & \colhead{description} & \colhead{salinity $S$ (g kg$^{-1}$)} &  \colhead{diffusivity $\kappa_{h}$ (m$^2$ s$^{-1}$)}
}
\startdata
\textit{HS$_{h}$} & High Salinity + high $\kappa_{h}$ & 35 & 0.5 \\
\textit{HS$_{l}$} & High Salinity + low $\kappa_{h}$ & 35 & 0.05 \\
\textit{LS$_{h}$} & Low Salinity + high $\kappa_{h}$ & 8.5 & 0.5 \\
\textit{LS$_{l}$} & Low Salinity + low $\kappa_{h}$ & 8.5 & 0.05 \\
\enddata
\end{deluxetable*}

\section{Experimental design}\label{sec:model}

We perform numerical simulations using the MITgcm to solve the non-hydrostatic equations for a Boussinesq fluid in a rotating spherical shell in a 3-D domain, where all sphericity terms are preserved, including all components of the Coriolis force \citep{adcroft2018mitgcm}. Radius, rotation rate and gravity are set to be the same as Enceladus, and the vertical variation of gravity is also taken into consideration \citep[see Table~\ref{tb:parameters} and][]{zeng2021ocean}. We use a non-linear equation of state \citep{jackett1995minimal} so that the thermal expansivity becomes negative near the freezing point in a low salinity ocean.

We simulate an ocean with a depth of 40~km on average over a zonal range of 15$^\circ$ with zonally periodic boundary conditions and a meridional range from 85.5$^\circ$S to 85.5$^\circ$N, with free-slip, no-normal flow conditions at the meridional boundaries. At the surface of the ocean, we prescribe the ice geometry and simulate the ice-ocean thermodynamics. In order to include the ice pump effect, we set the ice geometry to be an idealized (zonally-invariant and interhemispherically symmetric) approximation to Enceladus's inferred ice topography \citep{tajeddine2017true,vcadek2019long}, with ice thinning towards the pole. We do not allow the ice shell thickness to evolve (which would occur on a timescale that is much longer than the duration of the simulations) but compare the freezing and melting rate calculated in the simulations with the expected rate that is required to maintain the ice geometry against the ice flow. This approach is the same as used in \cite{kang2022symmetry} but different from \cite{kang2022saltice} (where the freezing and melting rates are prescribed). At the bottom of the ocean we apply a prescribed heat flux based on \cite{choblet2017powering} (Fig.~\ref{figS0_Setup}). The detailed model description can be found in APPENDIX~\ref{appendix:model}.

\begin{figure}[t]
\centering
\includegraphics[width=0.9\linewidth]{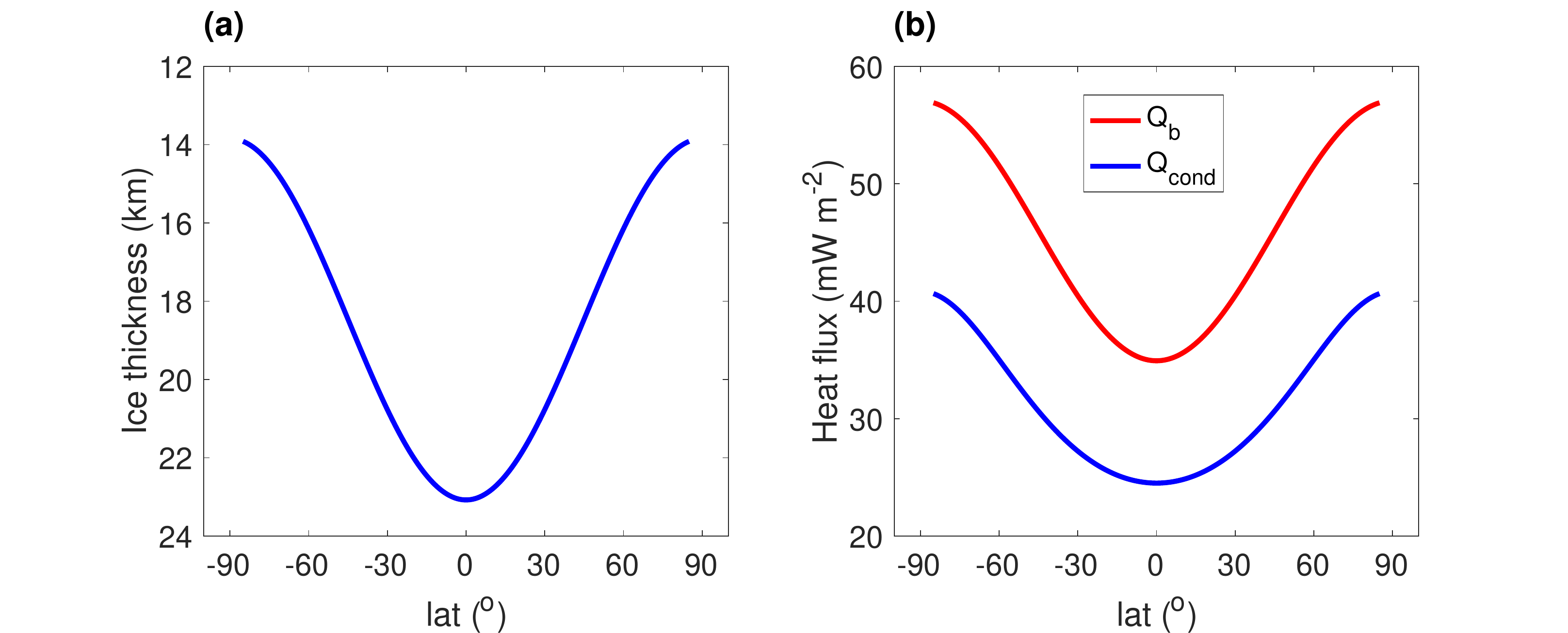}
\caption{Prescribed ice topography (a), corresponding conductive heat loss (b, blue line), and bottom heating (b, red line) per unit area, as a function of latitude. The global integral of bottom heating balances the conductive heat loss. The apparent offset between the blue and red lines in panel (b) is due to the different surface areas at the bottom and ice-ocean interface.}
\label{figS0_Setup}
\end{figure}

We carry out simulations with both high salinity (35~g~kg$^{-1}$, same as Earth ocean) and low salinity \citep[8.5~g~kg$^{-1}$, suggested by][]{glein2018geochemistry} to study the effect of salinity on ocean circulation and ice-ocean interaction on Enceladus. For each salinity, we use two different horizontal turbulent diffusivities ($\kappa_h$), 0.05~m$^2$~s$^{-1}$ and 0.5~m$^2$~s$^{-1}$, to test how sensitive the results are to this poorly constrained parameter, which is meant to mimic the effect of unresolved eddies and directly affects the meridional heat transport. The experimental setups are summarized in Table~\ref{tb:experiments}. All simulations are integrated to a quasi-equilibrium state where the total energy imbalance is less than 3\% of the seafloor heating rate (see APPENDIX~\ref{appendixsub:integration}).

\section{Simulation results}\label{sec:results}

\begin{figure}[b]
\centering
\includegraphics[width=1.0\linewidth]{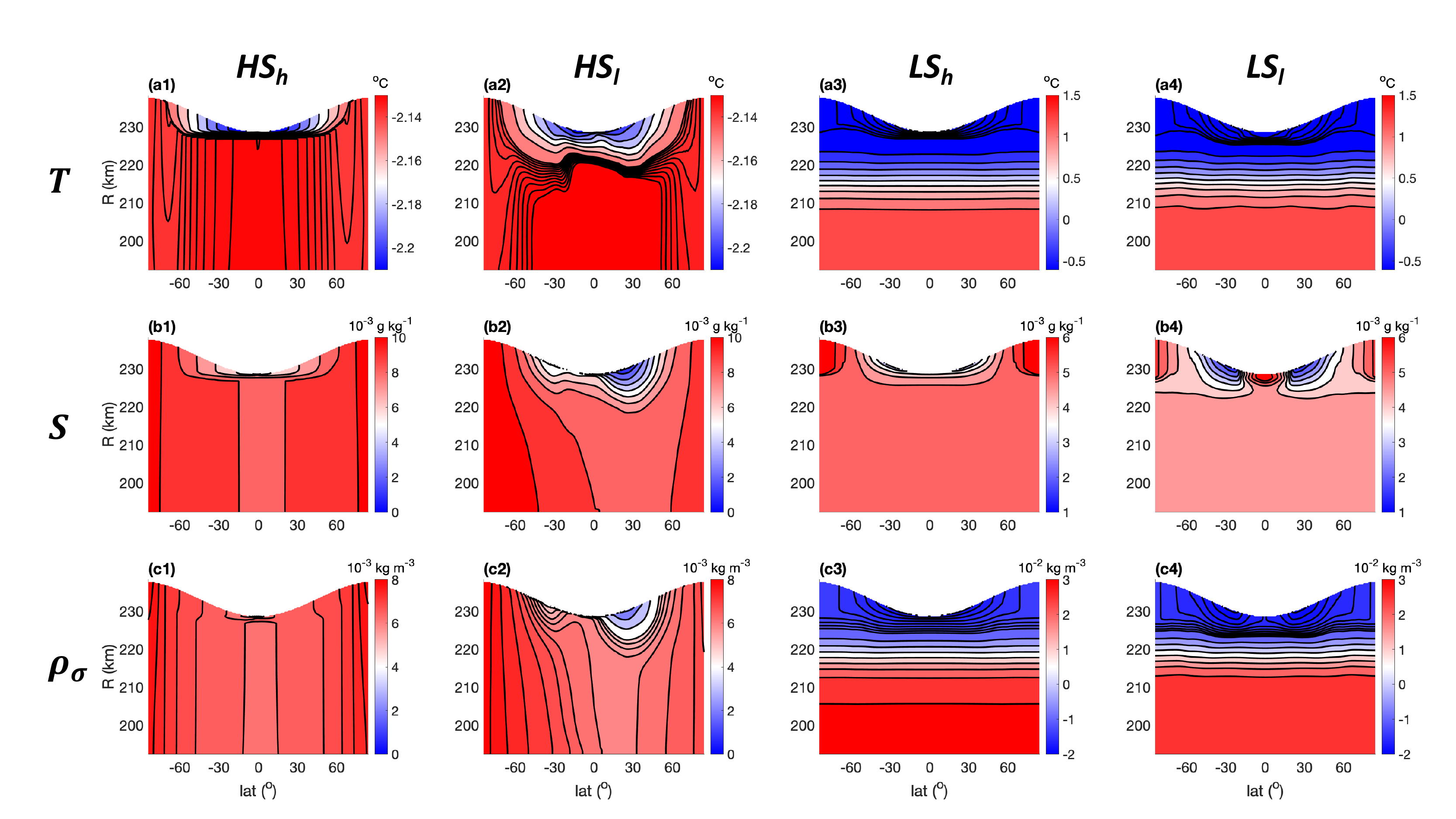}
\caption{Zonal-mean temperature ($T$, a), salinity anomaly ($S$, b), and potential density anomaly ($\rho_\sigma$, with reference pressure near the ocean surface, c) in the quasi-equilibrium state of all simulations. The first to the fourth columns are results of the high salinity simulation with high horizontal diffusivity (\textit{HS$_{h}$}), high salinity simulation with low horizontal diffusivity (\textit{HS$_{l}$}), low salinity simulation with high horizontal diffusivity (\textit{LS$_{h}$}), and low salinity simulation with low horizontal diffusivity (\textit{LS$_{l}$}), respectively. In (a1) \& (a2), the contour interval is 0.01~$^\circ$C below -2.14~$^\circ$C and 0.001~$^\circ$C above -2.14~$^\circ$C. In (a3) \& (a4), the contour interval is 0.01~$^\circ$C below -0.6~$^\circ$C and 0.2~$^\circ$C above -0.6~$^\circ$C. In (b1) \& (b2), a reference salinity of 34.97~g~kg$^{-1}$ is subtracted, and the contour interval is $10^{-3}$~g~kg$^{-1}$. In (b3) \& (b4), a reference salinity of 8.45~g~kg$^{-1}$ is subtracted, and the contour interval is $5 \times 10^{-4}$~g~kg$^{-1}$. In (c1) \& (c2), a reference density of 1029.13~kg~m$^{-3}$ is subtracted, and the contour interval is $10^{-3}$~kg~m$^{-3}$ below $5\times 10^{-3}$~kg~m$^{-3}$ and $3\times 10^{-4}$~kg~m$^{-3}$ above $5\times 10^{-3}$~kg~m$^{-3}$. In (c3) \& (c4), a reference density of 1007.7~kg~m$^{-3}$ is subtracted, and the contour interval is $10^{-3}$~kg~m$^{-3}$ below -$10^{-2}$~kg~m$^{-3}$ and $5\times 10^{-3}$~kg~m$^{-3}$ above -$10^{-2}$~kg~m$^{-3}$.}
\label{fig1_TSR}
\end{figure}

Ocean stratification is highly dependent on the mean ocean salinity. The high salinity ocean is mostly unstratified or weakly stratified below the ice-ocean interface. Instead, as previously found in ocean-only simulations with a flat upper boundary and constant salinity, a stably stratified layer develops in the low salinity ocean (Fig.~\ref{fig1_TSR}). In the stratified layer, the temperature decreases with height due to bottom heating, which, due to the negative thermal expansivity, amounts to a statically stable stratification \citep{zeng2021ocean}. The stratified layer maintains a large vertical thermal contrast due to the difference between the freezing point and the critical temperature at which the thermal expansivity changes sign (around 2~$^\circ$C difference with a salinity of 8.5~g~kg$^{-1}$, Fig.~\ref{fig1_TSR}a3 \& a4). Despite the ice shell topography, isopycnals in the stratified layer are largely flat rather than following the ice shell topography, indicating that ocean circulation is efficient in removing any large horizontal density gradients that would be associated with an undulating stratified layer.

\begin{figure}[b]
\centering
\includegraphics[width=1.0\linewidth]{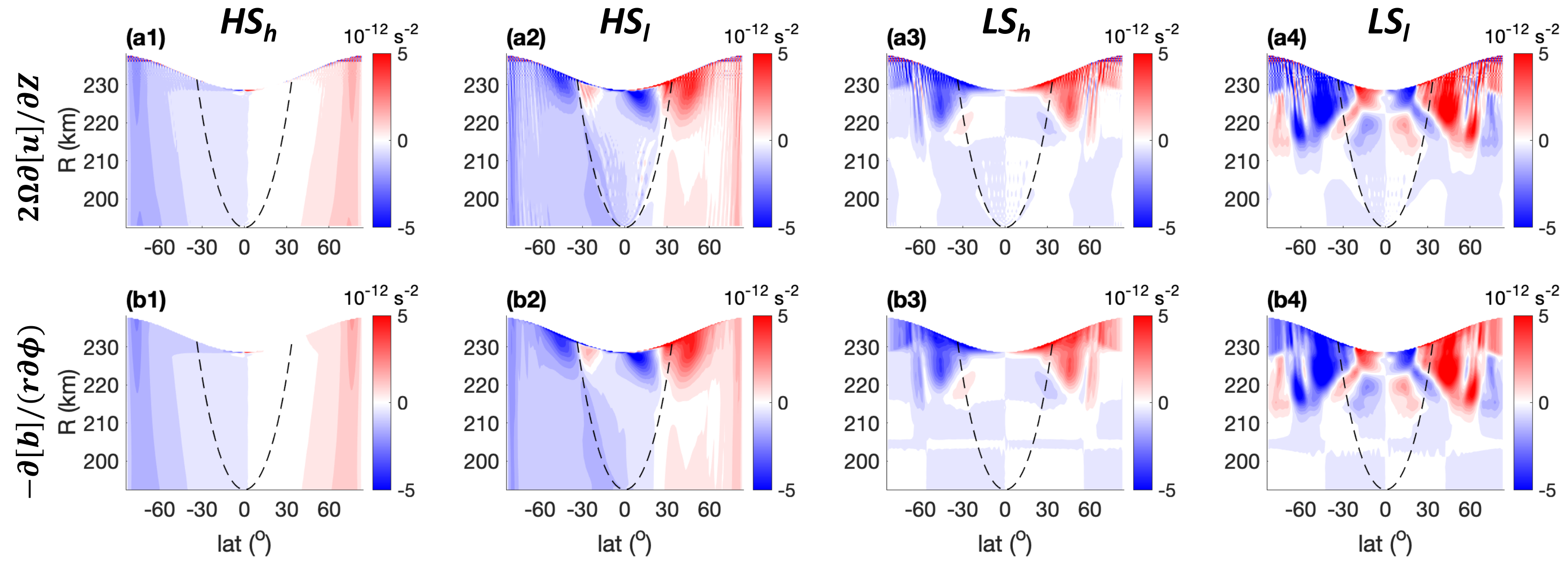}
\caption{Thermal wind balance in the quasi-equilibrium state of all simulations. (a) shows $2\vec{\Omega} \cdot \nabla [u]$, and (b) shows $-\partial [b]/r \partial \phi$, respectively, where $[\cdot]$ denotes a zonal average. The first to the fourth columns are results of the high salinity simulation with high horizontal diffusivity (\textit{HS$_{h}$}), high salinity simulation with low horizontal diffusivity (\textit{HS$_{l}$}), low salinity simulation with high horizontal diffusivity (\textit{LS$_{h}$}), and low salinity simulation with low horizontal diffusivity (\textit{LS$_{l}$}), respectively.}
\label{figS1_ThermalWind}
\end{figure}

\begin{figure}[t]
\centering
\includegraphics[width=1.0\linewidth]{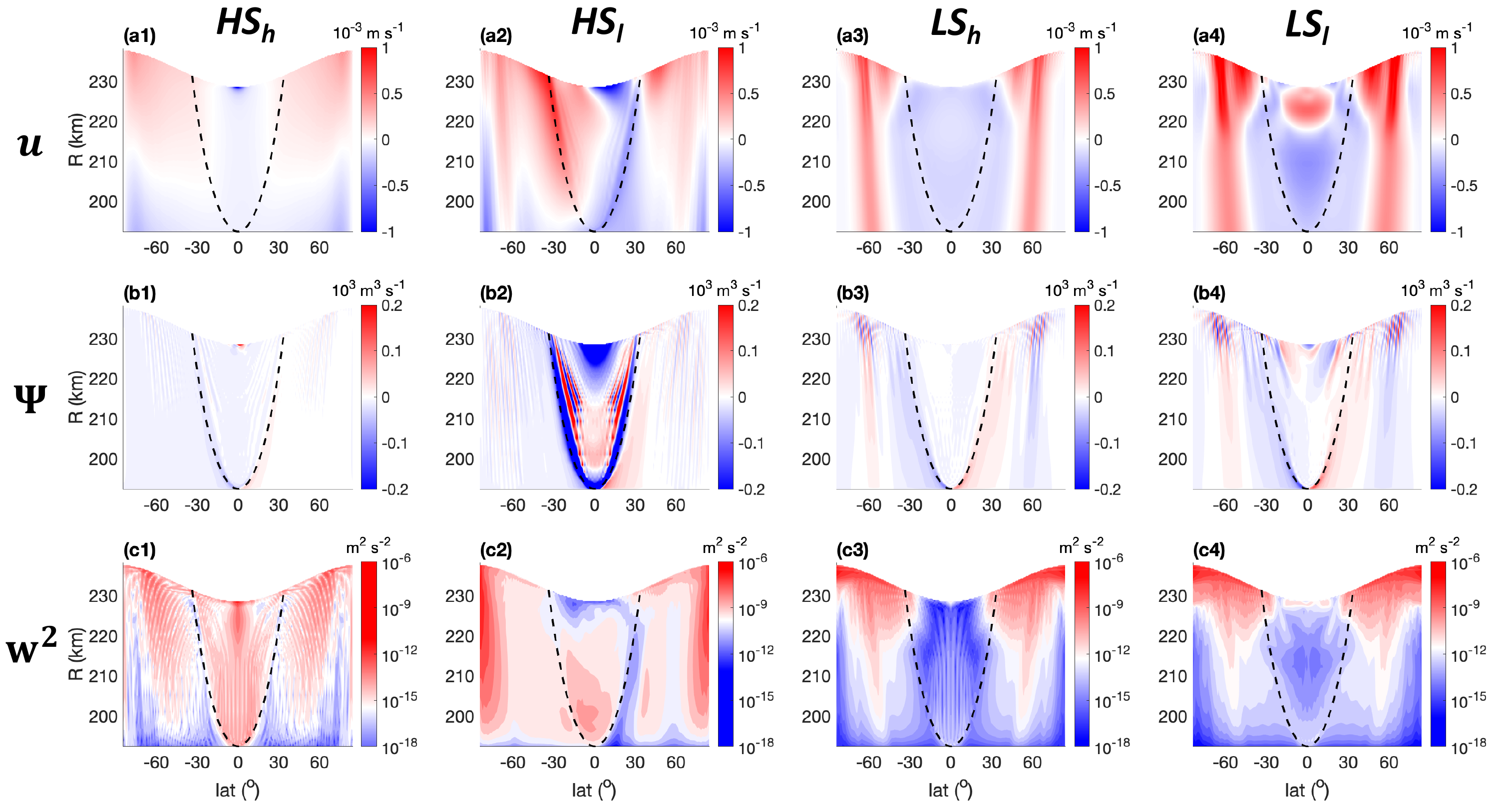}
\caption{Zonal-mean zonal velocity ($u$, a), stream function ($\Psi$, with clockwise circulation defined as positive, b), and zonal-mean vertical kinetic energy ($w^2$, in logarithmic scale, c) in the quasi-equilibrium state of all simulations. The first to the fourth columns are the results of the high salinity simulation with high horizontal diffusivity (\textit{HS$_{h}$}), high salinity simulation with low horizontal diffusivity (\textit{HS$_{l}$}), low salinity simulation with high horizontal diffusivity (\textit{LS$_{h}$}), and low salinity simulation with low horizontal diffusivity (\textit{LS$_{l}$}), respectively. The stream function is calculated as $\Psi = -\int_0^{\lambda_x} \int_{r_b}^r v r \cos{\phi} d\lambda dr$ where $\lambda_x=15\pi/180$ is the longitudinal extent of the domain, $v$ is the meridional velocity, $\lambda$ is longitude, and $r$ is radius. The black dashed lines denote the boundary of the tangent cylinder (a cylinder that is parallel to the rotational axis and is tangent to the bottom core).}
\label{fig2_UPW}
\end{figure}

The temperature and salinity structure is strongly affected by the surface ice shell in both high and low salinity simulations. In the upper ocean, the ocean is in contact with the ice shell, and the temperature is near the freezing point at the ice-ocean interface, which decreases with increasing pressure and hence with increasing ice thickness. Therefore, the temperature increases poleward in the upper ocean (Fig.~\ref{fig1_TSR}a). The salinity mostly increases poleward except for the equatorial region in the low salinity ocean (Fig.~\ref{fig1_TSR}b). The density mostly increases poleward in all simulations, although the role that temperature and salinity play in setting this structure differs in the high versus low salinity oceans (Fig.~\ref{fig1_TSR}c). In the high salinity ocean, the salinity and temperature gradients act in the opposite way, but the contribution from the salinity anomaly is about 2-3 times larger than that of the temperature anomaly, so that the density increases poleward. In the low salinity ocean, the thermal expansivity is negative, so that both temperature and salinity anomalies contribute to the poleward-increasing density, with the contribution from the temperature anomaly 3-5 times larger than the salinity anomaly. 

The density structure controls the zonal flow via the thermal wind relationship, which can be written as \citep[see e.g.][]{kaspi2009deep,bire2022exploring}:

\begin{equation}\label{eq:ThermalWind}
    2\vec{\Omega} \cdot \nabla u = 2 \Omega \frac{\partial u}{\partial Z} = 2 \Omega \left( \sin{\phi} \frac{\partial u}{\partial r} + \cos{\phi} \frac{\partial u}{r\partial \phi} \right) = - \frac{1}{r} \frac{\partial b}{\partial \phi},
\end{equation}

\noindent where $\vec{\Omega}$ is the planetary rotation rate, $u$ is the zonal velocity, $Z$ is the axis parallel to rotation, $r$ is radius, $\phi$ is latitude, and $b\equiv -g \delta \rho/\rho_0$ is buoyancy, where $g$ is gravity, $\delta \rho$ is the density anomaly and $\rho_0$ is a constant reference density. The thermal wind balance holds very well in all simulations (Fig.~\ref{figS1_ThermalWind}), except in regions where there is strong grid-scale convection (c.f. Fig.~\ref{fig2_UPW}b~\&~c). With density increasing poleward (i.e. $-\partial b/\partial \phi>0$ in the northern hemisphere; and opposite in the southern hemisphere), at mid- and high-latitudes we therefore find $\partial u/\partial Z > 0$, and there are eastward jets in the upper ocean with the jet cores aligned with the axis of rotation. Near the equator, $\sin{\phi}$ is close to zero and therefore $\cos{\phi} \partial u/(r\partial \phi)>0$, so that the zonal velocity decreases equatorward and there is a westward current near the equator (Fig.~\ref{fig2_UPW}a1-a3). An exception is that in the low salinity simulation with small horizontal diffusivity (\textit{LS$_{l}$}), an eastward flow exists near the equatorial surface (Fig.~\ref{fig2_UPW}a4) as a result of the locally reversed density gradient (Fig.~\ref{fig1_TSR}b4~\&~c4).

In both high and low salinity oceans, there is a slantwise Hadley-like circulation that ascends from the seafloor at the equator, following the tangent cylinder (a surface of constant angular momentum), and descends at higher latitudes. In the low salinity ocean, there is another reversed circulation cell at mid-latitudes along with an eastward jet, which bears similarity to the Ferrel cell in Earth's atmosphere. We also see grid-scale slantwise convection near the surface at high latitudes (Fig.~\ref{fig2_UPW}b). Although the radial stratification (parallel to the gravity vector) is near neutral or stable as maintained by the convective adjustment scheme in the model, the stratification along surfaces of constant angular momentum (mostly parallel to the rotational axis) can be unstable, which can cause slantwise convection.

\begin{figure}[t]
\centering
\includegraphics[width=1.0\linewidth]{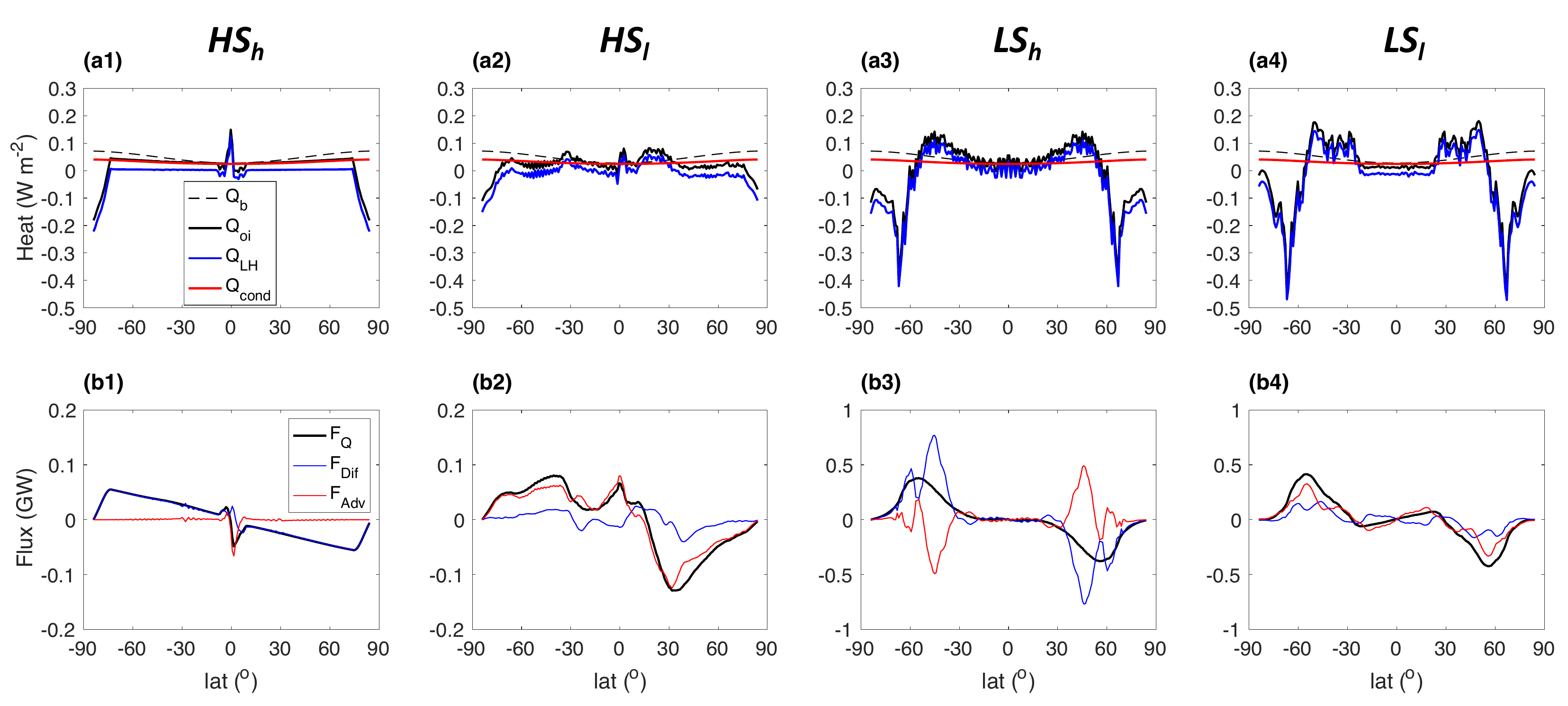}
\caption{Zonal-mean surface heat flux (a) and zonally- and vertically-integrated meridional ocean heat transport (b, see APPENDIX~\ref{appendix:oht} for details) in the quasi-equilibrium state of all simulations. The first to the fourth columns are results of the high salinity simulation with high horizontal diffusivity (\textit{HS$_{h}$}), high salinity simulation with low horizontal diffusivity (\textit{HS$_{l}$}), low salinity simulation with high horizontal diffusivity (\textit{LS$_{h}$}), and low salinity simulation with low horizontal diffusivity (\textit{LS$_{l}$}), respectively. In (a), the black dashed line is the bottom heating ($Q_{b}$), the blue line is the latent heating ($Q_{LH}$), the red line is the conductive heat loss ($Q_{cond}$), and the black solid line is the surface ocean-to-ice heat flux ($Q_{oi}$). In (b), the blue line is the diffusive heat flux ($F_{Dif}$), the red line is the advective heat flux ($F_{Adv}$), and the black line is the total heat flux ($F_{Q}$).}
\label{fig3_Heat}
\end{figure}

A strong small-scale overturning cell exists near the equatorial surface in the high salinity simulation with high horizontal diffusivity (\textit{HS$_{h}$}, Fig.~\ref{fig2_UPW}b1), which is associated with a positive feedback loop. There is a strong but narrow (only several degrees in latitude) melting region at and just south of the equator (Fig.~\ref{fig3_Heat}a1). As a result, a local density minimum and sharp density gradient exists at the equatorial surface (Fig.~\ref{fig1_TSR}c1). Following the thermal wind relationship, this density gradient is associated with a strong westward jet near the equatorial surface (Fig.~\ref{fig2_UPW}a1), and the corresponding frictional drag is associated with an Ekman transport that drives a localized overturning circulation. The local circulation cell then provides heat to maintain strong melting, forming a positive feedback loop. Note that this overturning is associated with a slight asymmetry around the equator where the peak of the melting and hence the salinity minimum is slightly displaced towards the southern hemisphere.

An even stronger asymmetry exists in the high salinity simulation with low horizontal diffusivity (\textit{HS$_{l}$}). There are overturning cells across the equator (Fig.~\ref{fig2_UPW}b2), and the salinity profile is asymmetric between the two hemispheres as a result of asymmetric freezing and melting patterns (Fig.~\ref{fig1_TSR}b2). The advection of heat and associated freezing and melting generates a positive feedback loop that can result in spontaneous symmetry breaking between the two hemispheres: once we get more melting in one hemisphere, the density becomes smaller, driving an overturning circulation across the equator, with sinking in the hemisphere with higher density and rising in the hemisphere with lower density. Due to the bottom heating, the temperature increases with depth at low latitudes, so that the overturning transports warmer water to the hemisphere with lower density, enhancing melting in that hemisphere. Note that in our simulation, this positive feedback seems to be confined to outside the tangent cylinder at low latitudes, although the details of the results are likely affected by poorly constrained model parameters and insufficient resolution. There is almost no asymmetry in the low salinity simulations, likely because the density gradient is dominated by the temperature gradient in the low salinity ocean. The temperature gradient is largely determined by the surface freezing point, which in turn is determined by the ice shell topography. As long as the ice shell topography remains symmetric (as assumed in this study) the temperature field also stays approximately symmetric (Fig.~\ref{fig1_TSR}a).

The ocean circulation drives ocean heat transport and modulates the surface density structure through the freezing and melting rate, and the density structure in turn modulates the ocean circulation. The heat budget at the ice-ocean interface is $Q_{cond} + Q_{LH} = Q_{oi}$, where the conductive heat loss ($Q_{cond}$) in the simulation is prescribed along with the ice geometry (see APPENDIX~\ref{appendixsub:boundary}). If the ocean is in an equilibrium state, the ocean heat flux to the ice shell ($Q_{oi}$) is balanced by the bottom heating ($Q_b$) when globally integrated. However, locally, the ocean heat transport determines the latent heat (i.e., the freezing and melting rate) by modulating the bottom heating pattern and providing an ocean heat flux to the ice shell.

\begin{figure}[ht]
\centering
\includegraphics[width=1.0\linewidth]{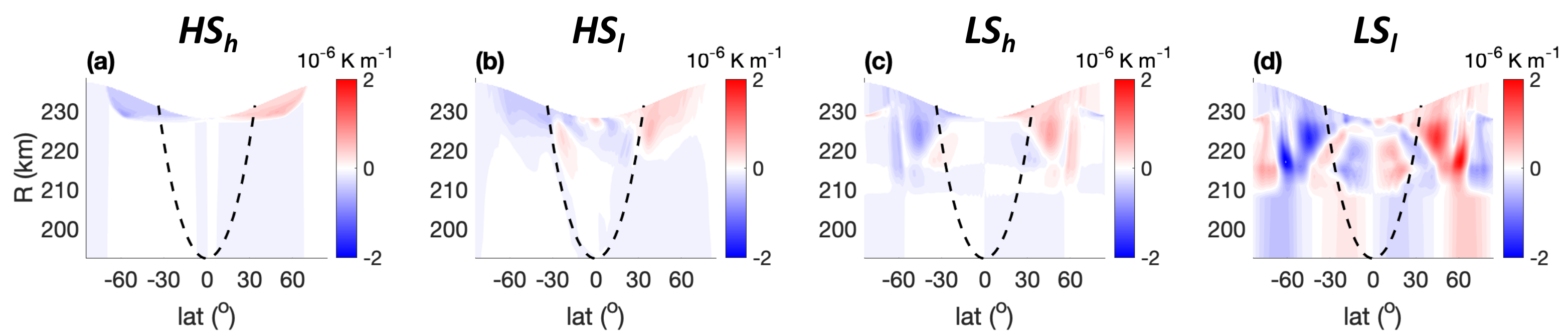}
\caption{Meridional temperature gradient $\partial T/(r\partial \phi)$ in the quasi-equilibrium state of all simulations. The first to the fourth columns are results of the high salinity simulation with high horizontal diffusivity (\textit{HS$_{h}$}), high salinity simulation with low horizontal diffusivity (\textit{HS$_{l}$}), low salinity simulation with high horizontal diffusivity (\textit{LS$_{h}$}), and low salinity simulation with low horizontal diffusivity (\textit{LS$_{l}$}), respectively. The black dashed lines indicate the tangent cylinder.}
\label{figS3_Ty}
\end{figure}

In all simulations, there is strong freezing at the pole and melting at lower latitudes (low latitudes in high salinity simulations and mid-latitudes in low salinity simulations; see Fig.~\ref{fig3_Heat}a). Locally, the dominant balance is between the latent heat flux ($Q_{LH}$) and the sensible heat flux from the ocean to the ice ($Q_{oi}$), because the spatial variations in the meridional heat flux convergence are much larger than the variations in the bottom heating and the conductive heat loss (Fig.~\ref{fig3_Heat}b). This latent heating profile, where freezing happens at higher latitudes than melting, is a result of the ocean heat transport mostly towards lower latitudes, induced by the ice pump effect. The meridional temperature gradient is strongest in the upper ocean where the ocean is in contact with the ice shell, so that the temperature gradient is largely determined by the freezing point at the ice-ocean interface (Fig.~\ref{figS3_Ty}). With polar-thinning ice geometry, the freezing point increases poleward so that the temperature is generally warmer at the pole than at the equator. As a result, any down-gradient heat flux is transporting warm water towards lower latitudes, so that melting occurs at lower latitudes than freezing. The strong freezing at the pole is moreover associated with a positive feedback: freezing at the pole leads to a density increase due to salt rejection, which destabilizes the stratification, thus promoting convection. Because the temperature decreases with depth under the ice shell at the poles (due to horizontal mixing with colder low-latitude water), the convection brings up colder water from below, which supports further freezing (see APPENDIX~\ref{appendix:surfaceheat}).

\begin{figure}[b]
\centering
\includegraphics[width=0.9\linewidth]{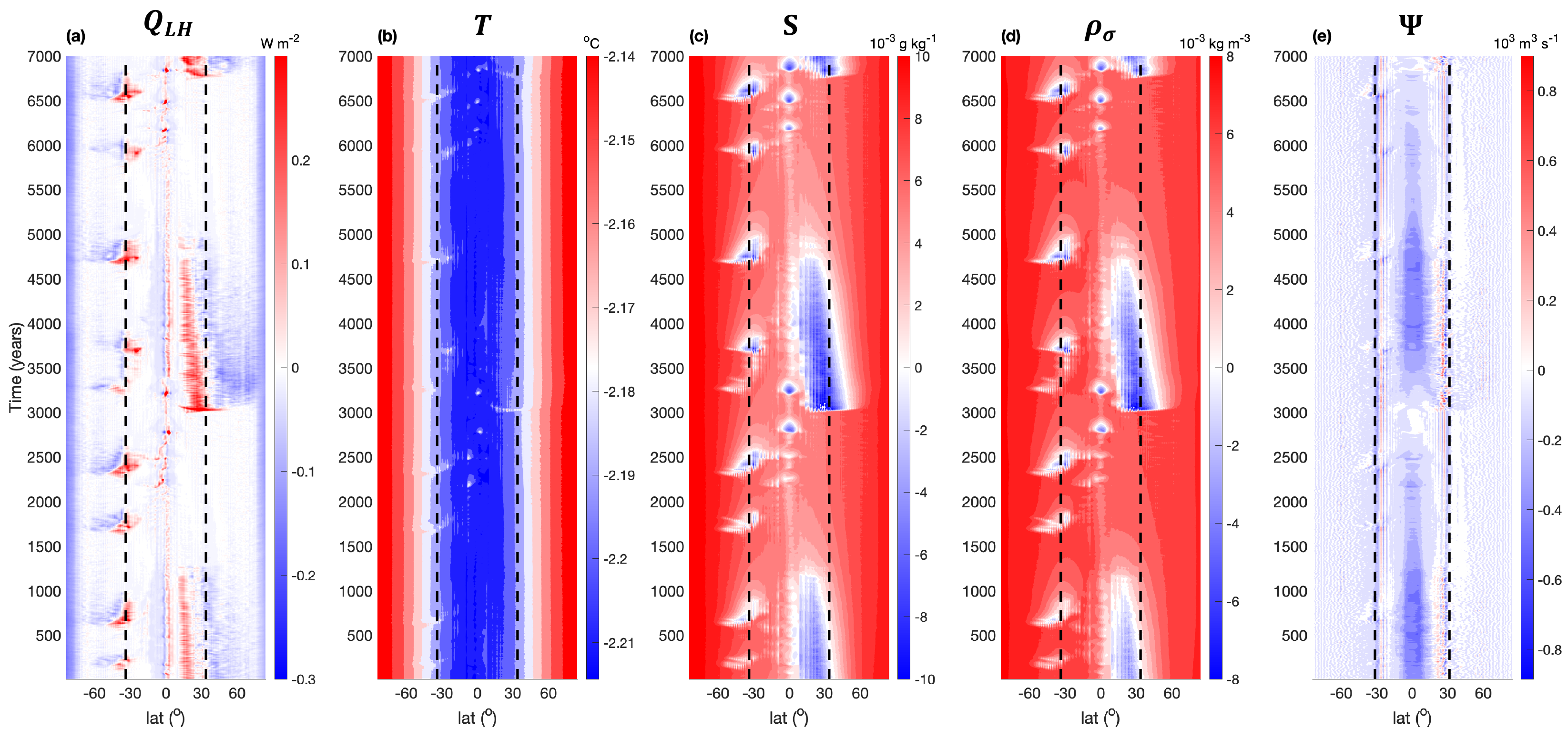}
\caption{Temporal variability in the high salinity simulation with low horizontal diffusivity (\textit{HS$_l$}). Zonal-mean latent heating (a), zonal-mean surface temperature (b), zonal-mean surface salinity (c, a reference salinity of 34.97~g~kg$^{-1}$ is subtracted), zonal-mean surface density (d, a reference density of 1029.13~kg~m$^{-3}$ is subtracted), and stream function at $r=225.375$~km (e) as a function of time and latitude. The black dashed lines are the latitudes where the tangent cylinder intersects the ocean surface in (a)-(d) and the latitudes where the tangent cylinder intersects $r=225.375$~km in (e). The first 4000-year average is used as the quasi-equilibrium state of the simulation \textit{HS$_l$}.}
\label{figS2_Variability}
\end{figure}

\begin{figure}[ht]
\centering
\includegraphics[width=0.9\linewidth]{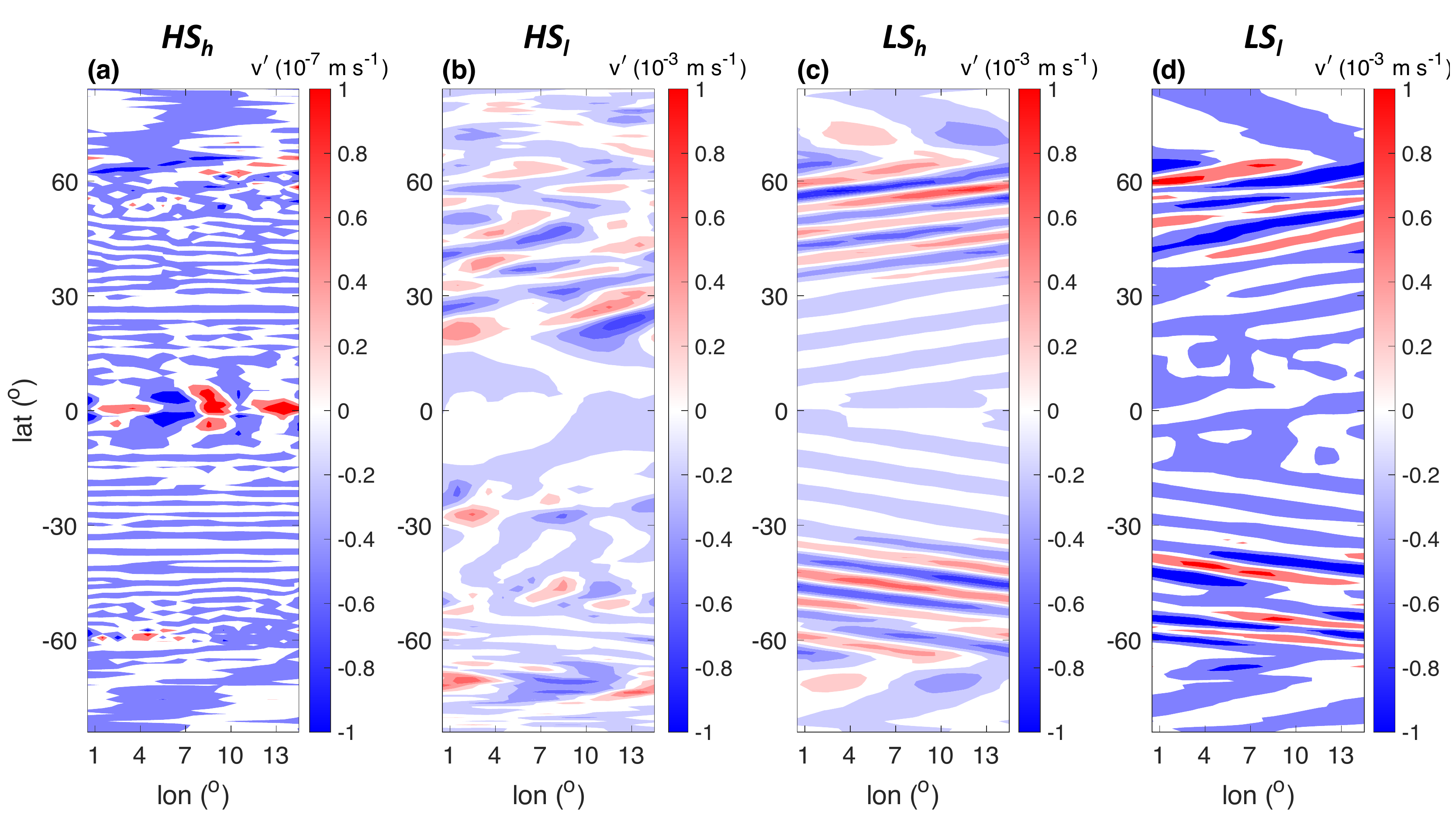}
\caption{Snapshots of meridional eddy velocity $v'$ (deviations from time- and zonal-mean) in all simulations at $r=225.375$~km. The first to the fourth columns are results of the high salinity simulation with high horizontal diffusivity (\textit{HS$_{h}$}), high salinity simulation with low horizontal diffusivity (\textit{HS$_{l}$}), low salinity simulation with high horizontal diffusivity (\textit{LS$_{h}$}), and low salinity simulation with low horizontal diffusivity (\textit{LS$_{l}$}), respectively. Note that the magnitude of $v'$ in (a) is $10^4$ times smaller than in (b-d).}
\label{figS4_EddyV}
\end{figure}

The meridional heat flux is larger in the low salinity ocean than in the high salinity ocean (Fig.~\ref{fig3_Heat}b). This result can be understood by noting that the meridional heat flux has three components: diffusive heat flux, resolved eddy heat flux, and heat flux by the mean circulation (See APPENDIX~\ref{appendix:oht}). All three fluxes become larger when there is a stratified layer. The meridional heat transport by diffusion and resolved eddies both increase with the vertical integral of the meridional temperature gradient. The meridional temperature gradient near the surface is similar in both high and low salinity oceans, because the temperature here is controlled by the freezing point, and hence is determined by the ice geometry, which is the same in both simulations. The meridional temperature gradient in the unstratified convective layer is small in all simulations. However, in the low salinity ocean, the temperature gradient in the stratified layer is comparable to that near the surface, so that the vertical integral of the meridional temperature gradient (i.e. the vertical extent of the meridional temperature gradient) is larger in the low salinity ocean (Fig.~\ref{figS3_Ty}), resulting in a higher meridional heat flux for a given eddy diffusivity. Additionally, the heat transport by the mean circulation is proportional to the strength of the overturning multiplied by the temperature contrast between the poleward and equatorward branches of the overturning circulation. Since the vertical temperature contrast is much larger in the low salinity ocean with a stratified layer, the heat transport by the mean circulation is also expected to be larger in the low salinity ocean.

Comparing simulations with different horizontal diffusivities indicates that, although freezing at the pole is found robustly across all simulations, the detailed pattern of melting is sensitive to model parameters. In the high salinity simulations, there is persistent melting at the equator in the simulation with higher diffusivity (\textit{HS$_{h}$}), while in the simulation with lower diffusivity (\textit{HS$_{l}$}), the melting latitude varies temporally. In the northern hemisphere, there is an intermittent melting region outside the tangent cylinder (at low latitudes), associated with the spontaneous symmetry breaking mechanism (Fig.~\ref{figS2_Variability}). The system switches between an asymmetric state and a more symmetric state with a period of about 1000-2000~years, but the cross-equatorial overturning never reverses in this simulation (although the sign of the circulation depends on the initial conditions – see APPENDIX~\ref{appendix:asymmetry}). In the southern hemisphere, there is intermittent melting in a region inside the tangent cylinder (at mid-latitudes). The southern hemisphere melting only persists for several hundreds of years and is likely associated with an instability near the tangent cylinder (Fig.~\ref{figS2_Variability}). The strong freezing at the pole and the pattern of melting in the low salinity simulations are less sensitive to horizontal diffusivities. In both high- and low-salinity ocean simulations, when the horizontal diffusivity $\kappa_h$ is high, the meridional heat transport is dominated by diffusion; when $\kappa_h$ is low, the advective heat transport becomes more important, while the overall heat flux changes relatively little (Fig.~\ref{fig3_Heat}b). The results indicate that increased eddy heat transport, associated with increased eddy activity (Fig.~\ref{figS4_EddyV}), tends to compensate for the decrease in the prescribed parameterized diffusivity. 

\section{Discussion and conclusion}\label{sec:conclusion}

We find that ice topography and salinity fundamentally affect the ocean stratification, circulation, and heat transport on Enceladus. In both high and low salinity scenarios, the upper ocean temperature, salinity and density all increase poleward. The temperature is near the freezing point at the ice-ocean interface, which varies with pressure and is thus set by the ice shell thickness variation. The salinity pattern is strongly affected by the salinity flux through freezing and melting. The density anomalies are dominated by the salinity anomalies in the high salinity simulations and are dominated by the temperature anomalies in the low salinity simulations, which is qualitatively consistent with \cite{kang2022saltice}.

In a low salinity ocean, with tidal heating at the bottom, a stably stratified layer forms in the upper ocean, as previously found by \cite{zeng2021ocean} in simulations with a flat upper boundary and no thermodynamic ice-ocean interface. This global thermally-driven stratified layer is caused by bottom heating and the negative thermal expansivity near the freezing point in a low salinity ocean, and is thus different from locally stable salt-stratified layers that form by surface melting, like the ``freshwater lens'' at the polar surface ocean discussed in \cite{lobo2020pole}, and the stratification in the ``shell-heating'' scenario in \cite{kang2022symmetry,kang2022saltice}. Note that the depth of the stratified layer here is deeper than in the more idealized simulations of \cite{zeng2021ocean}, but is thinner than in the ``core-heating'', low salinity ocean in \cite{kang2022symmetry,kang2022saltice}, where the entire ocean is stratified. The depth of the stratified layer $H$ is proportional to the vertical temperature contrast $|\Delta T_r|$, vertical diffusivity $\kappa_r$, and the inverse of the vertical diffusive heat flux $Q_{diff}$: $H \propto \kappa_r |\Delta T_r|/Q_{diff}$ \citep{zeng2021ocean}. The vertical turbulent diffusivity in \cite{kang2022symmetry,kang2022saltice} is set 100 times larger than in our simulations and in \cite{zeng2021ocean}, such that the depth of the stratified layer exceeds the depth of the ocean in \cite{kang2022symmetry,kang2022saltice}, leaving the entire ocean stratified. The smaller difference between our simulations and those in \cite{zeng2021ocean} can be understood by noting that all heat flux in the stratified layer is diffusive in the simulations in \cite{zeng2021ocean}, while in the simulations presented in this manuscript, some vertical heat transport is accomplished by advection. Therefore, $Q_{diff}$ here is somewhat smaller and $H$ is somewhat larger than in \cite{zeng2021ocean}. This small discrepancy is likely due to the much larger meridional temperature gradient in this work, which in turn arises from the freezing point variations due to the ice shell topography.

In our simulations, the convergence of the meridional heat flux is much stronger than the variation of the bottom heating and surface conductive heat loss. As a result, the ocean heat transport becomes the primary control on the pattern of ice freezing and melting, which is consistent with the core-heating scenario in \cite{kang2022symmetry} but is different from \cite{zeng2021ocean}, which uses a flat ice shell. With ice thickness decreasing from equator to pole, the freezing point increases poleward, resulting in a much stronger meridional temperature gradient and hence a stronger meridional heat flux in this work and \cite{kang2022symmetry} compared to \cite{zeng2021ocean} where the ice shell is flat.

The meridional heat flux is mostly towards lower latitudes in both high- and low-salinity oceans, due to the ice pump effect. This heat transport can be understood qualitatively by noting that the freezing point increases poleward due to the thinning of the ice shell, and the ocean heat transport tends to be directed down the meridional temperature gradient. Consequently, we find freezing at the poles, which cannot maintain the ice geometry against the poleward ice flow, as also found in \cite{kang2022symmetry,kang2022saltice}. The result that freezing happens at the poles indicates that either the ice shell on Enceladus is not in an equilibrium state, or that tidal heating in the ice shell is important in maintaining the ice geometry.

The stratified layer in the low salinity ocean tends to enhance the meridional heat transport by the mean overturning circulation through increasing the vertical temperature contrast, and enhance the meridional heat transport by diffusion and resolved eddies through increasing the vertical extent of the meridional temperature gradient. The first mechanism is also found in \cite{kang2022saltice}. In their core-heating scenario, the ocean heat transport is stronger in the low salinity ocean compared to the high salinity ocean, despite a weaker overturning.

In our high salinity simulation with small diffusivity, we find a positive feedback where enhanced melting in one hemisphere drives a cross-equatorial overturning circulation that further enhances the melting. This feedback can result in spontaneous symmetry breaking between the two hemispheres. Although the specific simulation results here cannot explain the observed ice thickness distribution, it is possible that a similar mechanism could play a role in the interhemispheric asymmetry of Enceladus's ice shell thickness. Indeed, a similar mechanism was previously identified by \cite{kang2022symmetry}. In the core-heating scenario in \cite{kang2022symmetry}, they however find that this ocean circulation acts to weaken the asymmetry of the ice shell, because melting occurs where the ice is thick in their simulations, which use an asymmetric ice shell topography. The combined results therefore indicate that the positive feedback may only hold as long as the asymmetry remains relatively small. With asymmetric tidal heating in the ice shell, \cite{kang2022symmetry} instead find enhanced melting in the hemisphere where the ice shell is thin, such that the ocean heat transport feedback can further enhance the asymmetry of the ice shell.

Besides Enceladus, previous studies have also investigated the heat transport in oceans on other icy moons \citep[e.g.,][]{soderlund2019ocean,amit2020cooling,ashkenazy2021dynamic,bire2022exploring,kvorka2022numerical}. In studies with direct numerical simulation (DNS) methods, simulations have been carried out without parameterizations of eddies and convection, and without including the effect of ice topography and ice-ocean thermodynamics \citep[e.g.,][]{soderlund2019ocean,amit2020cooling,bire2022exploring,kvorka2022numerical}. These studies reveal how the ocean circulation and ocean heat transport vary over a wide non-dimensional parameter regime. However, care must be taken when interpreting the real icy moon ocean regime with these simulations because the non-dimensional parameters in the numerical simulations and real icy moon oceans differ by many orders of magnitudes due to numerical constraints \citep{jansen2023energetics}. More directly comparable to the present work are studies using General Circulation Models (GCMs), where the effect of sub-grid-scale eddies (and in some cases convection) are parameterized \citep[e.g.,][]{ashkenazy2021dynamic,kang2022symmetry}. In \cite{ashkenazy2021dynamic}, the authors use MITgcm to study the ocean circulation and heat transport on Europa considering the effect of salinity and ice-ocean thermodynamics with a flat ice shell. They find that the meridional ocean heat transport is sufficient to compensate for the local imbalance between bottom heating and surface heat loss, thus allowing for a uniform ice thickness. This result is qualitatively consistent with our finding that ocean heat transport will tend to remove any ice thickness variations. However, the ice geometry is flat in the simulations of \cite{ashkenazy2021dynamic} so that there is no ice pump effect, and the planetary parameters of Europa and Enceladus are also significantly different, which makes it difficult to quantitatively compare the results of \cite{ashkenazy2021dynamic} to our work.

In global GCM ocean simulations of Enceladus and other icy moons, there are many processes that are difficult, or computationally impossible, to resolve. For example, slantwise convection cannot be adequately resolved but arises in the form of grid-scale convection in our simulations. Additionally, the effects of tides and librations are not explicitly included in our model - we only consider their effect in providing energy for eddy mixing in the choice for the vertical eddy diffusivity \citep[see Section 2.2 in][]{zeng2021ocean}. Therefore, we need parameterizations for these processes, which introduce external parameters (e.g., eddy diffusivties and viscosities) that are poorly constrained. Our results indicate that at least some of the results are sensitive to these parameterizations. The development of improved parameterizations for these unresolved processes in global icy moon ocean simulations therefore should be a high priority for future work.

\begin{acknowledgments}
We thank Wanying Kang, Edwin S. Kite, and two anonymous reviewers for helpful discussions and comments. This work was completed with resources provided by the University of Chicago Research Computing Center.
\end{acknowledgments}

\appendix

\section{Model description}\label{appendix:model}

\subsection{Parameterization of sub-grid scale processes}\label{appendixsub:kv}

In our simulations, the vertical resolution is 300-400~m near the boundaries, about 1000~m in the upper ocean, where the stratified layer is formed in a low salinity ocean, and 5000~m in the deep ocean. The varying vertical resolution is designed to save computational resources and to better resolve ice-ocean boundaries and the stratified layer in the low salinity ocean. The horizontal resolution is 1$^\circ$ in the zonal and 0.95$^\circ$ in the meridional direction, which is around 4~km~$\times$~4~km near the equatorial surface, decreasing to around 300~m~$\times$~3~km near the seafloor at the highest latitudes. 

Sub-grid scale processes are parameterized by turbulent viscosities and diffusivities. However, the turbulent viscosities and diffusivities are poorly constrained in simulations for ocean circulation on Enceladus. The vertical turbulent diffusivity in a stably stratified ocean can be constrained by the energy required to mix the water column \citep{wunsch2004vertical}. On Enceladus, this energy can come from tidal dissipation in the ocean and/or librations, based on which the vertical turbulent diffusivity in the stratified layer is estimated to be anywhere below $3\times 10^{-3}$~m$^2$~s$^{-1}$, although likely much smaller, according to \cite{zeng2021ocean}. Here we choose a vertical turbulent diffusivity of $5\times 10^{-5}$~m$^2$~s$^{-1}$ so that the estimated depth of the stratified layer is around 14~km and can be explicitly resolved. We further set the vertical viscosity to $\nu_r=5\times 10^{-4}$~m$^2$~s$^{-1}$ to keep the vertical Prandtl number $Pr=10$ \citep{soderlund2019ocean}. To ensure numerical stability, we apply a grid-dependent horizontal viscosity ($\nu_h$), increasing linearly with the square of the zonal grid scale ($L_x^2$). The horizontal viscosity maximizes at about 6~m$^2$~s$^{-1}$ near the equatorial surface and minimizes at about 0.5~m$^2$~s$^{-1}$ near the seafloor at the highest latitudes. We test two different horizontal diffusivities, 0.05~m$^2$~s$^{-1}$ and 0.5~m$^2$~s$^{-1}$. These values are motivated by \cite{kang2020core} who estimate the horizontal eddy diffusivity to be $\kappa_h = 0.8$~m$^2$~s$^{-1}$ using a scaling for geostrophically adjusted convection \citep{jones1993convection}.

The horizontal grid scale is not sufficient to resolve convective plumes. The length scale where rotation becomes important is estimated to be $l_r \approx 0.2$~m, many times smaller than our horizontal resolution, which means accurately resolving convective plumes is computationally impossible \citep{zeng2021ocean}. We therefore apply a convective adjustment by increasing the vertical diffusivity from the background value $\kappa_r=5\times 10^{-5}$~m$^2$~s$^{-1}$ to $\kappa_{conv}=$1~m$^2$~s$^{-1}$ whenever the stratification becomes unstable to parameterize the effect of convection. The value of $\kappa_{conv}$ is chosen such that the parameterized convective time scale ($\tau_{conv}=D_o^2/\kappa_{conv}\approx 50$~years where $D_o$ is the averaged depth of the ocean) is similar to the convective time scale estimated for the high salinity ocean \citep[see][]{zeng2021ocean}.

\subsection{Boundary conditions}\label{appendixsub:boundary}

We set the ice geometry based on Enceladus's topography and gravitational field as inferred by \cite{vcadek2019long}. We only apply the $Y_{20}$ component of the spherical harmonic function, which means we exclude zonal variations and hemispheric asymmetry in the ice geometry for simplicity  (Fig.~\ref{figS0_Setup}a).

At the bottom of the ocean, we apply a fixed bottom heat flux pattern as thermal boundary condition, following the zonal-mean tidal forcing pattern in \citet{choblet2017powering} with a total flux of 20~GW (Fig.~\ref{figS0_Setup}b). We are applying a smoothly varying bottom heating pattern, although there could be strong local hot spots in the core heating \citep[e.g.,][]{choblet2017powering}. While this may affect the details of the simulation results, it is not likely to affect the main conclusions. As recently shown by \cite{kang2022geyser}, warm water from a localized strong heat source will be quickly mixed with the surroundings due to baroclinic instability before it reaches the ice shell. As a result, the small-scale structure of the heating pattern is expected to have relatively little effect on the freezing and melting patterns at the ice-ocean interface. There is no salinity source or sink at the bottom of the ocean. We apply a linear bottom drag with a drag coefficient $r_b\sim 10^{-4}$~m~s$^{-1}$ as dynamic boundary condition (see \cite{zeng2021ocean} for a more detailed discussion of the bottom boundary condition). The same linear drag is also applied at the surface. The surface thermal and salinity boundary conditions are determined by the heat and salinity flux due to the ice-ocean interaction. 

We apply the ``shelfice'' package in MITgcm to calculate the ice-ocean interaction \citep{adcroft2018mitgcm}. The heat flux balance at the ice-ocean interface is between the conductive heat loss, the latent heat, and the ocean heat flux to the ice-ocean boundary. The conductive heat loss in the shelfice model is assumed proportional to the inverse of the ice shell thickness and the temperature difference between the freezing point and the ice surface temperature, the latter of which is assumed to be spatially constant in our simulations. The magnitude of the conductive heat loss is chosen such as to exactly balance the total heat flux at the sea floor, as is necessary to obtain an equilibrated ocean state. The ocean heat flux to the ice shell is calculated by a turbulent exchange flux through the ice-ocean boundary layer, where the heat flux is assumed proportional to the temperature contrast between the local ocean temperature and the temperature at the bottom of the ice shell (assumed to be the freezing point). The freezing and melting rate is calculated based on the latent heat required to balance the heat budget at the ice-ocean interface. In the shelfice model, the freezing and melting rate is converted to a salinity flux and does not affect the freshwater volume. The volume flux associated with meltwater input is negligible in our simulations: the latent heat is $Q_{LH} \sim$~O(0.1~W~m$^{-2}$) in all simulations (Fig.~\ref{fig3_Heat}a), so that the volume flux is $w_q=Q_{LH}/\rho L=3\times 10^{-10}$~m~s$^{-1}$ where $\rho$ is the density of the water and $L$ is the latent heat of fusion. This flux is negligible compared with the typical vertical velocity in the ocean ($w_o \sim$~O(10$^{-6}$~m~s$^{-1}$)). 

\begin{figure}[b]
\centering
\includegraphics[width=1.0\linewidth]{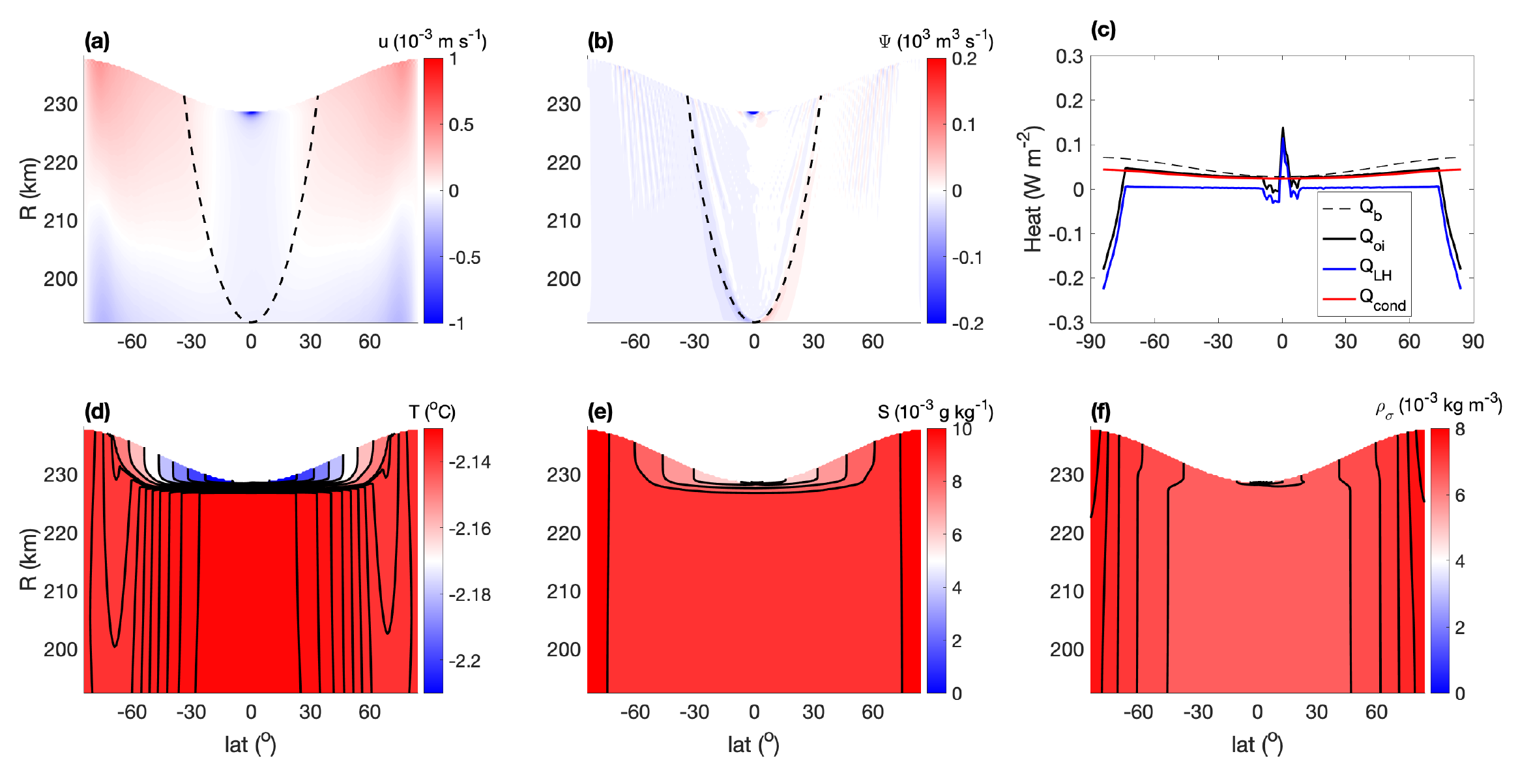}
\caption{Simulation with parameters in $HS_h$ but including surface temperature variation and temperature- dependent heat diffusion coefficient of the ice when prescribing the conductive heat loss \citep[following][]{kang2022saltice}. (a): zonal-mean zonal velocity $u$. (b): stream function $\Psi$, with clockwise circulation defined as positive. (c): surface heat budget, where the black dashed line is the bottom heating ($Q_b$), the blue line is the latent heating ($Q_{LH}$), the red line is the conductive heat loss ($Q_{cond}$), and the black solid line is the surface ocean-to-ice heat flux ($Q_{oi}$). (d): zonal-mean temperature $T$ (the contour interval is 0.01~$^\circ$C below -2.14~$^\circ$C and 0.001~$^\circ$C above -2.14~$^\circ$C). (e): zonal-mean salinity $S$ (a reference salinity of 34.97~g~kg$^{-1}$ is subtracted, and the contour interval is $10^{-3}$~g~kg$^{-1}$). (f): zonal-mean potential density $\rho_\sigma$ (a reference density of 1029.13~kg~m$^{-3}$ is subtracted, and the contour interval is $10^{-3}$~kg~m$^{-3}$ below $5\times 10^{-3}$~kg~m$^{-3}$ and $3\times 10^{-4}$~kg~m$^{-3}$ above $5\times 10^{-3}$~kg~m$^{-3}$).}
\label{figS5_RealSH}
\end{figure}

Our model does not account for the latitudinal variation of the ice surface temperature and the dependence of the heat diffusion coefficient of ice on the local ice temperature (both parameters are assumed constant in our model). We carried out one sensitivity test using the parameters of the simulation $HS_h$ but with the conductive heat loss through the ice shell computed using a latitudinally-varying surface temperature and a temperature-dependent heat diffusion coefficient, following \cite{kang2022saltice}. By comparing the results of this simulation with $HS_h$ in the main text, we find that the results do not qualitatively change (Fig.~\ref{figS5_RealSH}). Therefore, we do not expect these variations to qualitatively affect our results, as long as the conductive heat loss is increasing towards the pole, as a result of the polar-thinning ice shell.

It is not known whether Enceladus's ocean and ice shell are in an equilibrium state or not. The adjustment time scale of the ice shell can be estimated as $\tau_i \sim D_i/w_q \approx 2 \times 10^6$~years where $D_i=20$~km is the average ice thickness. For comparison, the (parameterized) convective time scale of the ocean is $\tau_{conv} \approx 50$~years, the overturning time scale of the ocean is $\tau_{adv} \sim V_o/ \Psi\approx 3 \times 10^5$~years where $V_o$ is the volume of the ocean and $\Psi$ is the stream function, and the diffusive time scale of the stratified layer in the low salinity ocean is $\tau_{diff} \sim H^2/\kappa_r \approx 10^5$~years, where $H$ is the thickness of the stratified layer and $\kappa_r$ is the radial diffusivity. We find that the adjustment time scale of the ice shell is much longer than the adjustment time scale of the ocean. As a result, although we do not know if the ice shell on Enceladus is equilibrated, it is more likely that the ocean is near an equilibrium state.

\subsection{Integration of the simulations}\label{appendixsub:integration}

The ice-ocean coupled 3-D global ocean simulations are difficult to integrate to a fully equilibrated state. Therefore, we present quasi-equilibrium states where the total energy imbalance is less than 3\% of the seafloor heating rate. Although there are still temperature and salinity trends in parts of the ocean, we expect that the remaining small imbalance does not qualitatively affect the general results discussed in this paper. The presented results are 500-year averages for \textit{HS$_{h}$}, \textit{LS$_{h}$}, and \textit{LS$_{l}$}, and 4000-year averages for \textit{HS$_{l}$} which has larger low-frequency variability.

We start the integration of all simulations with a uniform salinity (35~g~kg$^{-1}$ for high salinity and 8.5~g~kg$^{-1}$ for low salinity). The initial temperature for the high salinity ocean is set to be near the freezing point (about $-2^\circ$C) everywhere. For the low salinity ocean, we set the temperature in the upper ocean where the ocean is in contact with the ice shell to be near the freezing point (about $-0.6^\circ$C). At the bottom of the ocean, we set the temperature to the critical temperature where the thermal expansivity changes sign and becomes positive (about $1.4^\circ$C). In between, we calculate the depth of the stratified layer using Eq.~1 in \cite{zeng2021ocean} by assuming that the diffusive flux is equal to the total bottom heating, and then prescribe a linear temperature profile to connect the upper and bottom ocean layers, so that the initial conditions are reasonably close to the expected equilibrium state. Small-amplitude random noise is added to generate zonal variability.

This initial estimate still differs significantly from the final solution and the adjustment is very slow in the low salinity ocean. Therefore, we apply an acceleration method for the low salinity ocean simulations. For the temperature, we take the horizontally averaged profile for the temperature tendency ($\partial T/\partial t$), and accelerate the simulation by adding this tendency multiplied by a long time period (several thousands of years) to the original temperature profile. We only apply the tendency in layers where it has the same sign as the global temperature tendency: if the global ocean is net warming, we only apply positive temperature tendencies, and vice versa. After each acceleration step is applied, we integrate the model, initialized from the new profile, for about 200~years, and then repeat the acceleration step with the new tendencies. We repeat this procedure until the global temperature tendency changes sign, at which point we continue to integrate the model to a global thermal quasi-equilibrium state (defined as described above). It takes about 20 cycles (several tens of thousands of years) and a final integration of another several thousands of years for the low salinity ocean to reach a quasi-equilibrium state. During the acceleration steps, we did not accelerate the salinity, as the amplification of salinity tendencies from transient freezing and melting does not systematically nudge the simulated solution closer to its equilibrium state. Instead we apply a single adjustment step to the deep ocean salinity once a thermal quasi-equilibrium has been reached. In an equilibrium state, extrema of salinity can only exist at the surface of the ocean where there are salinity sources and sinks, so that every isoline of salinity should start or end at the surface. We hence find the lowermost isoline that intersects with the ice-ocean boundary and set the salinity below this isoline equal to this value. After this adjustment, we run the simulation for an additional 1000~years and use the last 500~years for analysis. Notice that the equilibrated solution is expected to be independent of the spin-up procedure and hence the acceleration method (unless multiple equilibria exist). While our low salinity simulation has not fully reached equilibrium, the remaining drifts are very small, suggesting that the results are not likely to differ qualitatively from the true equilibrium.

\subsection{Sensitivity to domain width}

\begin{figure}[b]
\centering
\includegraphics[width=1.0\linewidth]{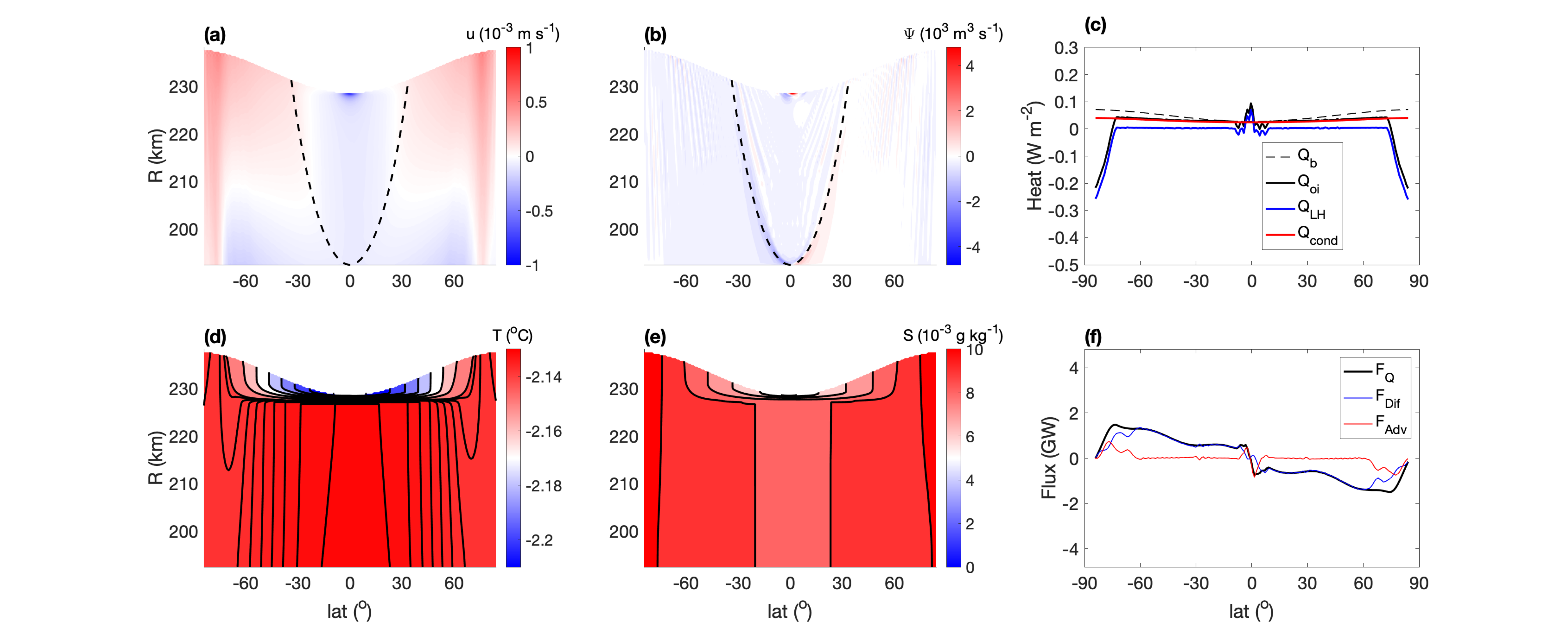}
\caption{Simulation with the same parameters as in $HS_h$ but with the zonal width of the domain extended from 15$^\circ$ to 360$^\circ$. (a): zonal-mean zonal velocity $u$. (b): stream function $\Psi$, with clockwise circulation defined as positive. (c): surface heat budget, where the black dashed line is the bottom heating ($Q_b$), the blue line is the latent heating ($Q_{LH}$), the red line is the conductive heat loss ($Q_{cond}$), and the black solid line is the surface ocean-to-ice heat flux ($Q_{oi}$). (d): zonal-mean temperature $T$ (the contour interval is 0.01~$^\circ$C below -2.14~$^\circ$C and 0.001~$^\circ$C above -2.14~$^\circ$C). (e): zonal-mean salinity $S$ (a reference salinity of 34.97~g~kg$^{-1}$ is subtracted, and the contour interval is $10^{-3}$~g~kg$^{-1}$). (f): zonally- and vertically-integrated meridional ocean heat transport. The blue line is the diffusive heat flux ($F_{Dif}$), the red line is the advective heat flux ($F_{Adv}$), and the black line is the total heat flux ($F_{Q}$). Note that the color bar range in (b) and the vertical axis range in (f) are multiplied by 24 compared to Fig.~\ref{fig2_UPW}b1 and Fig.~\ref{fig3_Heat}b1 to account for the larger zonal width of the domain and allow for easier comparison.}
\label{figS7_Global}
\end{figure}

In our simulations, we only simulate a zonal domain of 15$^\circ$ in longitude to save computational resources. However, the effect of horizontal eddies can be important in both ocean heat transport (see Fig.~\ref{fig3_Heat} and relevant discussions in the main text) and momentum transport \citep{ashkenazy2021dynamic}. Therefore, we carried out one sensitivity test with the same parameters as in $HS_h$, but with the zonal width of the domain extended from 15$^\circ$ to 360$^\circ$ (Fig.~\ref{figS7_Global}). To accelerate the simulation, we set the initial condition to be the zonal-mean field of the quasi-equilibrium state of $HS_h$ with small random perturbations added to the temperature field. This simulation has been run for 120~years, at which point a quasi-equilibrium state is reached where the total heat imbalance is smaller than 7\% of the seafloor heating rate. The results of this simulation are qualitatively similar to those in $HS_h$, in particular showing a very similar freezing and melting pattern.

\section{Heat budget}
\subsection{Meridional heat transport}\label{appendix:oht}

The meridional heat flux $F_Q$ has two components: advective heat flux $F_{Adv}$ and diffusive heat flux $F_{Dif}$. Here we define $F_Q$ as the zonally- and vertically-integrated, time-averaged heat flux, which is

\begin{equation}\label{eq4:heatflux}
    F_Q = F_{Adv} + F_{Dif} = \rho c_p \int_{0}^{\lambda_x} \int_{r_b}^{r_{oi}} \overline{vT} \ r \cos{\phi} d\lambda dr + \rho c_p \int_{0}^{\lambda_x} \int_{r_b}^{r_{oi}} \kappa_h \left( -\frac{1}{r} \frac{\partial \overline{T}}{\partial \phi}\right) \ r \cos{\phi} d\lambda dr,
\end{equation}

\noindent where $r_{b}$ is the radius at the bottom of the ocean, $r_{oi}$ is the radius at the ice-ocean interface, $v$ is the meridional velocity, $T$ is the temperature, $\lambda$ is longitude, $\lambda_x$ is the longitudinal extent of the domain, and the overbar indicates a time-average. The time- and zonally-averaged advection can be written as $ [\overline{vT}] =  [\overline{v}] [\overline{T}] +  [\overline{v'T'}]$, where the bracket denotes the zonal average and primes denote deviations from the zonal and temporal average - which we will refer to as the eddy component. Eq.~\ref{eq4:heatflux} can then be written as

\begin{equation}\label{eq5:average}
    F_Q = \rho c_p \lambda_x \int_{r_b}^{r_{oi}} [\overline{v}] [\overline{T}] \ r \cos{\phi} dr + \rho c_p \lambda_x \int_{r_b}^{r_{oi}} [\overline{v'T'}] \ r \cos{\phi} dr - \rho c_p \lambda_x \int_{r_b}^{r_{oi}} \kappa_h \frac{\partial [\overline{T}]}{\partial \phi} \ \cos{\phi} dr.
\end{equation}

If we assume that the eddy heat transport is down-gradient and is proportional to the mean gradient of temperature, we can apply an equivalent ``eddy diffusivity'' $\kappa_e$ to replace the eddy term: $[\overline{v'T'}] = -\kappa_e \partial [\overline{T}]/r\partial \phi$ \citep{vallis2017atmospheric}. Then Eq.~\ref{eq5:average} becomes

\begin{equation}\label{eq6:eddyflux}
    F_Q = \rho c_p \lambda_x \int_{r_b}^{r_{oi}} [\overline{v}] [\overline{T}] \ r \cos{\phi} dr - \rho c_p \lambda_x \int_{r_b}^{r_{oi}} (\kappa_e+\kappa_h) \frac{\partial [\overline{T}]}{\partial \phi} \ \cos{\phi} dr,
\end{equation}

\noindent where on the right hand side the first term indicates heat transport by the mean overturning circulation, while the second term indicates heat transport by resolved eddies and diffusion (which may be thought of as representing the effect of unresolved eddies).

\subsection{Ocean heat budget of the surface layer}\label{appendix:surfaceheat}

\begin{figure}[t]
\centering
\includegraphics[width=1.0\linewidth]{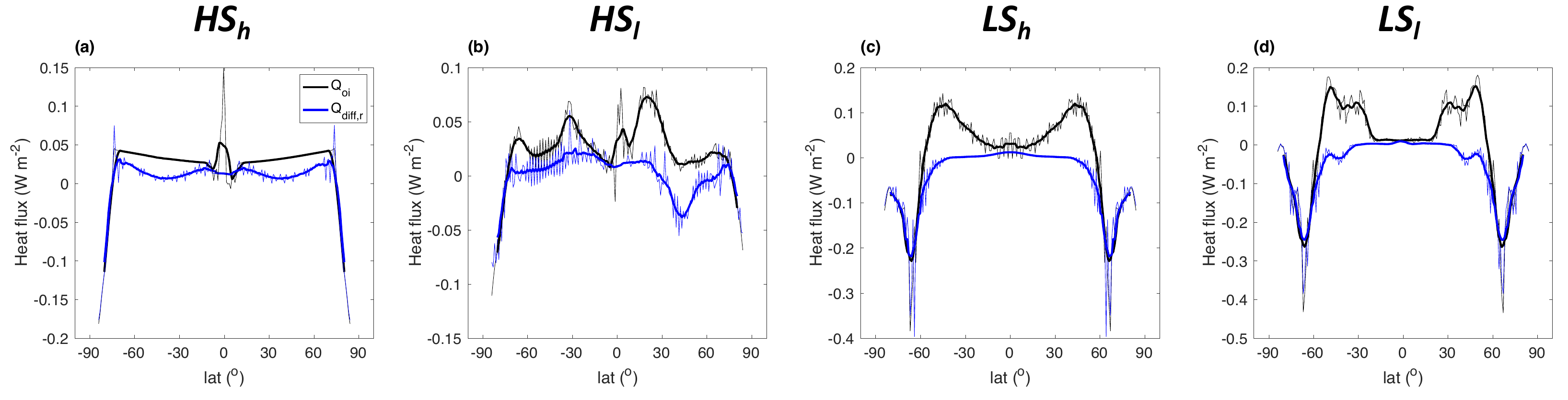}
\caption{Vertical heat fluxes near the surface of the ocean (the uppermost ocean model layer in contact with the ice shell). The blue lines show the radial diffusive heat flux (including the parameterized convective heat flux) at the bottom of the surface ocean model layer, and the black line shows the ocean heat flux to the ice shell ($Q_{oi}$). The thinner lines are the raw data and the thicker lines are moving-averages of 9 grids to smooth grid-scale variabilities. The first to the fourth columns are results of the high salinity simulation with high horizontal diffusivity (\textit{HS$_{h}$}), high salinity simulation with low horizontal diffusivity (\textit{HS$_{l}$}), low salinity simulation with high horizontal diffusivity (\textit{LS$_{h}$}), and low salinity simulation with low horizontal diffusivity (\textit{LS$_{l}$}), respectively.}
\label{figS8_SurfaceHeat}
\end{figure}

To better understand how the parameterized upright convection is connected with the surface freezing and melting pattern, we compare the vertical diffusive heat flux near the ice-ocean interface (at the bottom of the model layer in contact with the ice shell) with the ocean heat flux to the ice shell $Q_{oi}$ (Fig.~\ref{figS8_SurfaceHeat}). The parameterized vertical convective diffusivity ($\kappa_{conv}=1$~m$^2$~s$^{-1}$) is much larger than the background vertical diffusivity ($\kappa_{r}=5\times 10^{-5}$~m$^2$~s$^{-1}$), so that the vertical diffusive heat flux primarily shows the effect of the parameterized upright convection. In all simulations, near the polar surface, $Q_{oi}$ approximately matches the parameterized upright convective heat flux, indicating that the convection brings up colder water from below and supports freezing near the poles.

\begin{figure}[ht]
\centering
\includegraphics[width=1.0\linewidth]{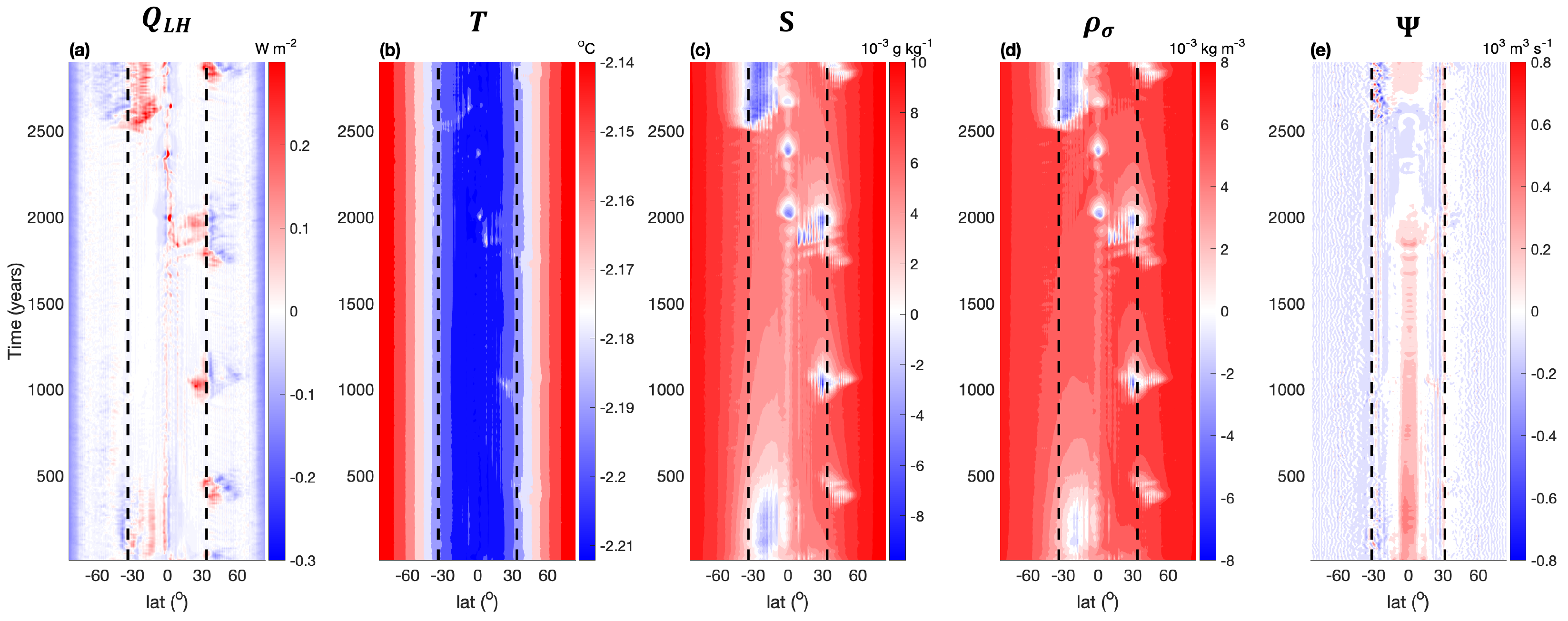}
\caption{Simulation with the same parameters as in $HS_l$, but using the ``mirrored'' salinity field of the quasi-equilibrium state in $HS_l$ as the initial condition. Zonal-mean latent heating (a), zonal-mean surface temperature (b), zonal-mean surface salinity (c, a reference salinity of 34.97~g~kg$^{-1}$ is subtracted), zonal-mean surface density (d, a reference density of 1029.13~kg~m$^{-3}$ is subtracted), and stream function at $r=225.375$~km (e) as a function of time and latitude. The black dashed lines are the latitudes where the tangent cylinder intersects the ocean surface in (a)-(d) and the latitudes where the tangent cylinder intersects $r=225.375$~km in (e).}
\label{figS6_Flip}
\end{figure}

\section{Multiple equilibria in high salinity simulations} \label{appendix:asymmetry}

We have found a positive feedback which can result in spontaneous symmetry breaking in the high salinity simulations. This mechanism, however, should not have a preference between the northern and southern hemisphere. We hence carried out one sensitivity test with the same parameters as the $HS_l$ simulation but using the “mirrored” salinity field of the quasi-equilibrium state in $HS_l$ as the initial condition. In this sensitivity test, we observe that the cross-equatorial overturning circulation has the opposite sign. Consistently, we find stronger melting outside the tangent cylinder at low latitudes in the southern hemisphere (in contrast to the northern hemisphere in $HS_l$), and we still see the switching between an asymmetric state and a more symmetric state (Fig.~\ref{figS6_Flip}). The nonlinear dynamics controlling the multiple quasi-equilibria could be an interesting topic for future studies.

\bibliography{Enceladus_OcnIce}{}
\bibliographystyle{aasjournal}

\end{document}